\documentclass[onecolumn,preprintnumbers,amsmath,amssymb]{revtex4}

\usepackage{graphicx}
\usepackage{dcolumn}
\usepackage{bm}

\begin{document}

\preprint{}

\title{On Daryl Bem's Feeling the Future Paper}

\author{Akhila Raman }

\affiliation{Email: akhila.raman@berkeley.edu}

\date{\today}

\begin{abstract}

It has been argued by Daryl Bem in his 2011 paper $^{[1]}$ that  8 out of 9 experiments yielded statistically significant results in favour of the psi effect.
It is pointed out in this short communication that many of the results in the above mentioned paper could be explained by using well known concepts in statistics such as Confidence Level and Standard Error of the Sample Mean.

\end{abstract}

\maketitle

\section{\label{sec:level1}Section 1\protect\\  \lowercase{} }

It has been argued by Daryl Bem in his 2011 paper $^{[1]}$ that  8 out of 9 experiments yielded statistically significant results in favour of the psi effect.
[retroactive influence by "time reversing" well-established psychological effects so that the individual's responses are obtained before the putatively causal stimulus events occur.]\\

Let us consider the first experiment "Precognitive Detection of Erotic Stimuli"
which had 100 participants with 36 trials each[ 40 sessions showing randomly displayed 12 erotic and 24 non-erotic pictures and 60 sessions displaying 18 erotic and 18 non-erotic pictures]. In each trial, the participants were asked to guess one of the 2 curtains behind which the picture would be displayed.  The statistically significant(0.05) hit rate was reported to be $53.1 \%$  for the case of erotic pictures. Here the sample size is $N=1560 (40*12+60*18)$. \\

Each trial in this experiment is a Bernoulli trial and can be modelled by a binary Random Variable(RV) $\textbf{X}_{i}$ which takes one of 2 values, 0 or 1, to indicate the subject's guess. $\textbf{X}_{i}$ is a Bernoulli RV and has a mean of $0.5$ and standard deviation $\sigma_b = 0.5$. Let us assume the null hypothesis that the population mean is $0.5$. \\

We then add all such  guesses for the whole sample size of $N=1560$ and then divide by $N$ to estimate the sample mean. Hence we get another random variable $\textbf{Y}= \frac{1}{N} \displaystyle\sum\limits_{i=1}^{N} \textbf{X}_{i}$ which has an approximately Gaussian distribution (Central Limit Theorem) with mean $\mu_g$ and a standard deviation$^{[2]}$ $\sigma_{g}=\frac{\sigma_b}{\sqrt{N}}$ which is called the \textbf{Standard Error of the Sample Mean} (SEM). \\

Now we can compute the confidence interval(limits) of this approximately Gaussian Distribution, for a specified Confidence Level.  Confidence Level is given by the equation $ P( \mu_g - n * \sigma_g < \textbf{Y} <  \mu_g + n * \sigma_g ) = erf(\frac{n}{\sqrt{2}}) $  and refers to the probability that $\textbf{Y}$ lies within the confidence interval (which is "n" times the standard deviation, computed from the mean). Corresponding to $95\%$ Confidence Level, confidence interval $CI =1.96*\sigma_g$ $^{[3]}$. Hence we can write the Confidence Interval as $CI =1.96*\frac{\sigma_b}{\sqrt{N}}$. Given that  $\sigma_b = 0.5$, we have $CI =\frac{0.98}{\sqrt{N}}$ for a $95\%$ Confidence Level. 
[We can also consider other values of Confidence Level like $99 \%$ and so on] \\

A $95\%$ Confidence Level implies that if we repeat this experiment $M$ times, as $M \to \infty$, for every 95 out of 100 such experiments, the sample mean would be within this Confidence Interval $CI =\frac{0.98}{\sqrt{N}}$. Which means that for 
every 5 out of 100 such experiments, the sample mean would be \textbf{outside} this Confidence Interval $CI =\frac{0.98}{\sqrt{N}}$ as $M \to \infty$. For finite $M$, for \textbf{approximately} 5 out of 100 such experiments, the sample mean would be \textbf{outside} this Confidence Interval $CI =\frac{0.98}{\sqrt{N}}$. \\

For the experiment under discussion, $N=1560$ and hence $CI =\frac{0.98}{\sqrt{1560}} = 0.0248$ and for approximately 5 out of 100 such experiments, sample mean would be $ > 0.5+0.0248 =0.5248$ which is close to observed hit rate of $0.531$. If we consider a $99\%$ Confidence Level, corresponding Confidence Interval $CI =2.5 * \frac{0.5}{\sqrt{1560}} = 0.0316$ [ $erf(\frac{2.5}{\sqrt{2}}) = 0.99 $] and for approximately 1 out of 100 such experiments, sample mean would be $ > 0.5+0.0316 =0.5316$ which is close to observed hit rate of $0.531$. Daryl Bem's experiment could very well have been one such experiment(1 out of 100 experiments).The reason why other groups have not succeeded in replicating Daryl Bem's results could be the fact that their experiments fell in the other 99 out of 100 experiments category. This argument could be applied to other 8 experiments by Daryl Bem.\\

If we repeat this experiment $M$ times where $M$ is large and average across the ensemble, the sample mean may get closer to the expected population mean of $0.5$.\\

Which means that when a certain experiment produces anomalous results with $0.05$ statistical significance, this means that the results correspond to  $95\%$ Confidence Level  and the anomalous results are to be expected for 5 out of 100 such experiments for which the sample mean would be \textbf{outside} this Confidence Interval $CI =\frac{0.98}{\sqrt{N}}$. If we repeat that experiment $M$ times where $M$ is large and average across the ensemble, the sample mean may become closer to the expected population mean of $0.5$. Or if the results are truly anomalous, the sample mean averaged across the ensemble will remain closer to the anomalous sample mean.\\

The same argument could be applied to opinion polls published by organizations like Gallup.

\section{\label{sec:level1}References\protect\\  \lowercase{} }

[1]  Bem, D.J. (2011). Feeling the Future: Experimental Evidence for Anomalous Retroactive Influences on Cognition and Affect. Journal of Personality and Social Psychology, 100(3), 407-425. 
[Online version:  http://dbem.ws/FeelingFuture.pdf  ]\\

[2] Athanasios Papoulis, Probability, random variables, and stochastic processes.
New York : McGraw-Hill, c1984. pp178-194.\\

[3] Confidence Level and Confidence Interval. Mathworld. http://mathworld.wolfram.com/ConfidenceInterval.html\\

\end{document}